\documentclass[11pt,a4paper]{article}
\pdfoutput=1
\usepackage{myjheppub}
\usepackage{latexsym,amsfonts,amsmath,amssymb}
\usepackage{mathtools}
\usepackage{bbm}
\usepackage{graphicx}
\usepackage{color}
\usepackage[font=small,labelfont=bf]{caption}
\usepackage{subcaption}
\usepackage[normalem]{ulem}
\usepackage{comment}
\usepackage{url}
\usepackage{slashed}
\usepackage{bm}

\newcommand{\CE}{\mathcal{E}}

\newcommand{\CN}{\mathcal{N}}
\newcommand{\CO}{\mathcal{O}}

\newcommand{\CH}{\mathcal{H}}

\newcommand{\CI}{\mathcal I}
\newcommand{\CQ}{\mathcal Q}

\renewcommand{\Re}{{\rm Re}}
\newcommand{\Tr}{\mbox{Tr}}

\newcommand{\IC}{\mathbb{C}}
\newcommand{\IZ}{\mathbb{Z}}

\newcommand{\rme}{{\rm e}}

\newcommand\be{\begin{equation}}
\newcommand\ee{\end{equation}}
\newcommand\bea{\begin{eqnarray}}
\newcommand\eea{\end{eqnarray}}

\renewcommand{\=}{\;= \;}
\newcommand{\+}{\;+ \;}

\renewcommand{\t}{\tau}

\newcommand{\wt}{\widetilde}

\renewcommand{\Re}{\text{Re}}

\renewcommand{\i}{{\rm i}}

\newcommand{\Seff}{S_\text{eff}}

\def\t{\tau}

\newcommand{\is}{i_\text{s}}
\newcommand{\ig}{i_\text{grav}}
\newcommand{\x}{{\rm x}}
\newcommand{\SBH}{S_\text{BH}}
\newcommand{\blue}{\textcolor{blue}}
\newcommand{\bl}{\pmb \lambda}
\newcommand{\defeq}{\; \coloneqq \;} 

\title{The growth of the $\frac{1}{16}$-BPS index in 4d $\CN=4$ SYM}

\author{Sameer Murthy}

\affiliation{Department of Mathematics, King's College London,\\
  The Strand, London WC2R 2LS, U.K}

\emailAdd{sameer.murthy@kcl.ac.uk}

\abstract{We study the Hamiltonian index of $\frac{1}{16}$-BPS operators in 4d~$\CN=4$ $U(N)$ 
super Yang-Mills (SYM) theory numerically for~$N=2\,,\dots,10$. 
We show that the large-charge asymptotics agree with analytic results in the Cardy-like limit, 
as consistent with the entropy of supersymmetric black hole in the dual AdS. 
The numerics also agree with the large-$N$ analytic result, thus providing hints towards an exact formula for the index. 
We then prove, using ideas from representation theory, that for 
values of charges (quantized in integer units) 
less than~$2(N+1)$ the index agrees precisely with the multi-graviton index, and then begins to deviate 
for larger charges. Thus the~$U(N)$ SYM index interpolates between multi-graviton values at small charge 
and black hole growth at large charges.
}

\begin{document}

\addtolength{\topmargin}{2cm}

\maketitle

\addtolength{\topmargin}{-2cm}


\section{Introduction and summary}

This note discusses
a certain counting problem regarding $\frac{1}{16}$-BPS operators in $\CN=4$~$U(N)$ SYM theory,
that is to say, operators annihilated by one supercharge~$\CQ$ of the theory and its Hermitian conjugate. 
Upon considering the theory on~$S^3 \times R^t$ and performing radial quantization, the problem can be 
recast as one regarding $\frac{1}{16}$-BPS states in the theory, and we will henceforth use these two ideas interchangeably. 
The problem is to verify if there is a gravitational interpretation for these BPS states consistent 
with the AdS/CFT correspondence.
In particular, one expects that at small values of charges the number of BPS states should equal that of an ensemble 
of supergravitons, while at large charges the statistical entropy of the ensemble of BPS states 
should agree with the thermodynamic entropy of $\frac{1}{16}$-BPS black holes in the dual~AdS$_5 \times$~S$^5$ theory. 
This problem was originally formulated and discussed in~\cite{Sundborg:1999ue, Aharony:2003sx, Kinney:2005ej}, and 
has led to a lot of interesting work since then. 
The conclusion of this note is that the~$\CN=4$ SYM index indeed exhibits the behavior expected from AdS/CFT in both limits, 
and interpolates between them as a function of charge.

\vskip 0.2cm

The context of the counting problem is as follows.
The gravitational theory has a scale set by the radius of the asymptotic~AdS$_5$ in five-dimensional Planck units, 
which corresponds to a positive power of the rank~$N$ of the gauge group. 
The BPS black hole solution is then specified by four independent charges---three~R-charges labelling
the representations of~$SU(4)_R$ and two angular momenta on~$S^3$, constrained by one relation. 
In the $\CN=4$ SYM theory, $\frac{1}{16}$-BPS states are labelled by the charges that commute with~$\CQ$,
these are the two angular momenta and two of the three R-charges.\footnote{The non-linear relation between
the five charges in the gravitational theory is not directly visible in the SYM theory, and this is part of what makes this 
problem subtle. Unravelling this issue was an important part of the recent 
progress~\cite{Hosseini:2017mds, Cabo-Bizet:2018ehj}.} 
The essential part of the problem can be formulated in any~$\CN=1$ SCFT with a gravity dual,
where one has a single~$R$-charge~$Q$ and the two angular momenta. 
In fact, in the simplest (and first to be discovered) 
BPS black hole solution~\cite{Gutowski:2004yv}, only one combination~$J$ of the two angular momenta is non-zero. 
This is the simplest setting within which one can study the basic problem of microscopic black hole entropy. 
In this setting, the supercharge~$\CQ$ can be chosen such that the combination of charges that commutes with it 
is~$2J+Q$. The quantization condition on~$J$ is universal, while the quantization of~$Q$ depends on the theory 
under consideration. 
For~$\CN=4$ SYM written in this~$\CN=1$ language the quantity~$n=3(2J+Q)$ is a positive integer,
this is the situation we discuss in this note.

\vskip 0.2cm

The phrase ``large/exponential growth of states", which is often used in this context, deserves a small discussion.
As just explained, the counting problem has two scales---the rank~$N$ and the charge~$n$. 
The regime of validity of the black hole solution is~$N \to \infty$ (classical  
gravity theory), and~$j \equiv n/N^2$ non-zero (large horizon area\footnote{The full non-BPS spectrum of the gravitational
theory also contains the so-called ``small" BH, which is really a ten-dimensional meta-stable solution. We do not have 
any comments to make about this here.}). The thermodynamic entropy of these black holes in supergravity 
is given by
\be \label{SBH}
\SBH \; \equiv \;  \frac{1}{4} \, A_\text{H}  \=  N^2 \, s(j) \,,
\ee
where~$s(j)$, a function which we discuss in some detail below, behaves as follows as~$j \to \infty$,
\be \label{sSBH}
s(j) \= \frac{\pi}{2 \cdot 3^{1/6}} \, j^{2/3}  + O(j^{1/3})\,.
\ee
The expectation of the exponential growth of states is the question of whether the logarithm of the 
indexed number of microscopic states reproduces the behavior~\eqref{sSBH} for large charges. 
More precisely, consider the superconformal index\footnote{The generating function~$\CI(\x) = \sum d(n) \, \x^n$ 
is also often called the superconformal index. To avoid confusion, we call~$\CI(\x)$ the index~\emph{function}.} 
$d_N(n)$ defined as the trace of~$(-1)^F$ over the subspace of the 
Hilbert space with charge~$n$ in the~$U(N)$ theory. We want to address whether its asymptotic behavior 
agrees with the thermodynamic entropy of the black hole, i.e.,
\be \label{dnasymp}
\frac{1}{N^2} \log d_N(n) \; \stackrel{?}{\longrightarrow}\;   s(n/N^2)  \quad \text{as} 
\quad N \to \infty \,, \quad j\=\frac{n}{N^2} \quad \text{fixed}. 
\ee
This is the~\emph{large-$N$ limit} relevant to our problem.

\vskip 0.2cm

Recent work~\cite{Hosseini:2017mds,Cabo-Bizet:2018ehj,Choi:2018hmj, Choi:2018vbz, Benini:2018ywd,
Honda:2019cio, ArabiArdehali:2019tdm,Zaffaroni:2019dhb,Kim:2019yrz,Cabo-Bizet:2019osg,Amariti:2019mgp,
Lezcano:2019pae,Lanir:2019abx,Cabo-Bizet:2019eaf,ArabiArdehali:2019orz}
has shown that the answer to the question posed in \eqref{dnasymp} is yes.
There have been essentially three different approaches, 
all of which begin by considering the index function\footnote{\label{qtaurel}Here the variable~$\x$ 
is related to~$q=\rme^{2 \pi \i \t}$ defined in~\cite{Cabo-Bizet:2019eaf} as~$q=\x^3$.}
\be \label{indtrace}
\CI_N (\x) \= {\rm Tr}_{\CH_\text{phys}}\,  (-1)^F \rme^{-\beta \{\mathcal{Q},\bar{\mathcal{Q}}\} } \, 
\x^n \= \sum_n d_N(n) \, \x^n \,,
\ee
to which only~$\CQ$-invariant states contribute and which is therefore independent of~$\beta$.
This index function can be calculated as an integral over~$N \times N$ unitary 
matrices~\cite{Romelsberger:2005eg}, \cite{Kinney:2005ej}, 
\be \label{Uact}
\CI_N(\x) \= \int \, DU\, \exp \biggl( \; \sum_{j=1}^\infty \, \frac{1}{j} \, 
\is(\x^j) \, \Tr \, U^j \, \Tr \, (U^\dagger)^j \, \biggr) \,,
\ee
where~$\is(\x)$ is the index trace as in~\eqref{indtrace} but taken over all single ``letters" of the 
gauge theory. 
The new approaches to this problem involve new 
techniques to calculate~$\CI_N(\x)$, or to estimate its behavior as~$|\x| \to 1$. 
Having obtained this, 
one can then invert the relation~\eqref{indtrace} to obtain~$d_N(n)$ using a 
saddle-point analysis, which is valid as~$n \to \infty$. 
Two of these approaches---the Bethe ansatz approach~\cite{Benini:2018ywd} and 
the direct saddle-point analysis of the matrix integral~\cite{Cabo-Bizet:2019eaf}---directly address the large-$N$ limit. 
In both these approaches, the supersymmetric index is governed by certain special functions---the elliptic gamma 
function~\cite{Felder, Spiridonov:2012ww} and the Bloch-Wigner elliptic dilogarithm~\cite{Bloch},  
respectively---whose properties help us calculate the asymptotics of the corresponding degeneracy of black holes 
in~AdS.

The third approach~\cite{Choi:2018hmj,Honda:2019cio, ArabiArdehali:2019tdm,Kim:2019yrz,
Cabo-Bizet:2019osg,Amariti:2019mgp} addresses a related but slightly different problem, 
namely the growth of~$d(n)$ as~$n \to \infty$ at \emph{fixed}~$N$, or, equivalently, $j \to \infty$ at finite~$N$, 
which is called the \emph{Cardy-like limit}. This limit corresponds to a black hole whose horizon radius is 
very large in AdS units or, more pictorially, ``fills up AdS space".
One can then take~$N \to \infty$ to obtain classical gravity in AdS space.
The results of these three approaches are consistent with each other in overlapping regimes of validity.

\vskip 0.2cm

In this note we study the matrix integral~\eqref{Uact} for finite values of~$N$.
Instead of relying on properties of special functions as in the analytic approaches, we want to 
gain an understanding from the point of view of the Hamiltonian trace. 
There are two aims of this study. Firstly we want to verify the recently-obtained analytic results 
quoted above and thus compare the numerical values of the microscopic index~$d_N(n)$ to the black hole entropy. 
Secondly we want to understand the nature of~$d_N(n)$ as a function of~$n$, and 
clarify some older discussion about the~$Q$-cohomology of~$\CN=4$ SYM, in particular 
the relation to the so-called ``graviton partition function"~\cite{Grant:2008sk,Kinney:2005ej,Chang:2013fba}.

Now, what do the above-mentioned limits mean on a computer?
For the Cardy-like limit we fix~$N$, and study the growth of~$d_N(n)$ as~$n$ becomes large.
In order to treat the large-$N$ limit, we fix~$j$ to be some small finite number, 
then count the index of states with~$n=j N^2$ for a given~$N$, 
and finally take a limit where we increase~$N$ and~$n$ simultaneously to large values.

\vskip 0.2cm

In Section~\ref{sec:large} we report on numerics for~$N=2,\dots,10$. 
The numerics show agreement with the analytic results in the Cardy-like limit.\footnote{Some numerical results
for this problem showing agreement with the Cardy limit were reported in~\cite{KimStrings, Choi:2018vbz}.}
In fact they serve to illustrate one further point of detail which is worth emphasizing . 
The results of~\cite{Kim:2019yrz,Cabo-Bizet:2019osg} show that the answer for the entropy in the Cardy-like limit for finite~$N$
agrees exactly with the supergravity calculation, including the full function~$s(j)$ 
and the~$N^2$ dependence as in~\eqref{SBH},~\eqref{sSBH}. 
A priori, this need not have been the case---one could have had a functional dependence on~$N$
which only reached~$N^2$ asymptotically as~$N \to \infty$ for finite charge. Indeed, our experiments 
with low values of~$N$ not only reflect the~$n^{3/2}+\dots$ behavior but in fact show 
beautiful agreement with the full function~\eqref{SBH}, \eqref{SBHsfunction},
including (surprisingly) at fairly small values of charges and~$N$. 
Perhaps this agreement is pointing towards a simple exact quantum entropy formula at finite~$N$ and finite charges.

\vskip 0.2cm

Our numerical results become more interesting when one contrasts the behavior of~$d_N(n)$ at 
small~$n$ with that at large~$n$. First we discuss large~$n$. 
Consider the Hamiltonian interpretation of the integral~\eqref{Uact}. 
The~$j=1$ term inside the exponential is the single-letter trace index, the sum over~$n$ and the exponential
operation lift this counting to the full multi-trace spectrum, and the integral over~$U$ implements the 
projection to gauge-invariant states. 
It is not difficult to show that the exponential of the single-letter trace by itself  
gives a $n^{2/3}$ charge dependence. This alone, however, can not reproduce the required scaling in~$N$
because the adjoint character~$\Tr U \, \Tr U^\dagger$ projected to gauge invariant states 
vanishes for~$SU(N)$ and gives a single state for~$U(N)$.
The interpretation of the large-$N$ analytic results combined with the Cardy-like asymptotics
mentioned above is that, for finite~$N$ and large charge~$n$,
roughly $N^{2/3}$ degrees of freedom of the matrix get deconfined so that, combined with the~$n^{2/3}=N^{4/3} j^{2/3}$ 
Cardy-like scaling, we obtain the~$N^2$ behavior in~\eqref{dnasymp}. 
The numerical and analytical agreement with the full supergravity entropy function explained in the 
previous paragraph are consistent with this. 

\vskip 0.2cm

In Section~\ref{sec:small} we look at the other regime~$n \ll N$. Here one has complete control 
over the problem from a different point of view, namely global symmetry. It is convenient to phrase 
this in the holographic language, but we should keep in mind that one only uses the global symmetry 
of supergravity on~AdS$_5 \times$~S$^5$ or, equivalently, that of~$\CN=4$ SYM. In this regime, 
using the fact that one can ignore all trace relations among matrices, it was shown in~\cite{Chang:2013fba} 
(based on earlier work in \cite{Janik:2007pm, Grant:2008sk}) how to construct the complete set of 
gauge-invariant operators of the SYM theory and that this 
set agrees exactly with the set of multi-graviton states in the gravitational theory. It was further conjectured 
in~\cite{Chang:2013fba} that the~$\CQ$-cohomology does not contain any other states even at finite~$N$. 

Here we first explain and verify the equality of the infinite-$N$ SYM states and the graviton states from 
the perspective of the matrix integral~\eqref{Uact}. The notable point about multi-gravitons is that they 
do not depend on the rank~$N$ at all, and only on the charge~$n$. We present an argument based on 
representation theory  that explains that the values of the index correspond precisely to the index of 
multi-gravitons for states with low-lying charge~$n \le 2N+1$.  For higher values of charge~$n$ new 
states do begin to contribute, so that as~$n$ reaches~$ j N^2$ there is an exponential growth of states.  
The index~$d_N(n)$ thus interpolates from the graviton behavior at small charge to the black hole behavior at large charge, 
which is the finite-$N$ microcanonical manifestation of the Hawking-Page transition.

\section{The index at infinite~$N$ and at finite~$N$ \label{indrev}}

In this section we first review the puzzle of the $\frac{1}{16}$-BPS states, then we review some of the 
old results about the index at infinite~$N$ and the relation to the multi-graviton states, then we review 
the Hamiltonian interpretation of the index at finite~$N$ as a matrix integral. 

\vskip 0.2cm

Firstly, why is the index expected to have an exponential growth of states equal to that of a black hole? 
The answer of course comes from AdS/CFT---the index counts a difference between the number of 
bosonic and fermionic $\frac{1}{16}$-BPS states, and so is a lower bound on the total number of such states
in the field theory. 
On the gravitational side, the existence of a black hole in the gravitational spectrum of the theory can be interpreted 
as there being a number of states equal to the exponential of the entropy of the black hole, and so the 
growth of states of the index should reflect that. This line of reasoning, of course, presumes a Hilbert space 
interpretation of the quantum gravitation theory on~AdS about which very little is directly known. 

The Euclidean version of the argument is more clear in this respect---the statement of the 
AdS/CFT conjecture in this context is that the functional integral of the boundary SYM theory should be equal 
to the functional integral of the AdS space with appropriate boundary conditions. 
In the field theory one interprets the index trace as a functional integral as usual. 
However, it is not clear a priori from this point of view whether the Euclidean BPS black hole solution in AdS$_5$ 
contributes to the AdS functional integral, even at the classical level where the functional integral 
can be approximated by the exponential of the on-shell action.
The issue is one of regulating the infra-red behavior of the BPS black hole  
solution. This problem was addressed in~\cite{Cabo-Bizet:2018ehj} where it was shown, by considering a supersymmetric
deformation of the BPS BH away from extremality, that there is a regulator consistent with 
supersymmetry using which the on-shell action equals the BPS BH entropy in the limit. 
One thus reaches the statement that the index should grow at least as fast as the 
exponential of the BPS BH entropy at large-$N$.\footnote{There are further possible corrections to this statement 
coming from (a) quantum corrections to the BH entropy, (b) the possible existence of other saddles in the AdS
functional integral, and (c) the possibility of wall-crossing when one flows from weak to strong coupling,
(see~\cite{Mandal:2010cj,Dabholkar:2012nd,Dabholkar:2011ec,Dabholkar:2014ema} for a discussion of 
these issues in BHs in asymptotically flat space), none of which we will discuss here.}
\footnote{It may be possible, along the lines of~\cite{Sen:2009vz,Dabholkar:2010rm}, to make a more precise statement 
about~\emph{equality} of the index and the black hole entropy, we will not discuss this here.}

\vskip 0.2cm

Next we discuss the issue of loop corrections. 
It is well-known that the $\frac{1}{16}$-BPS operators are not protected in the SYM theory because 
the Hamiltonian is corrected at one-loop~\cite{Beisert:2003jj}.
Thus the~$\CQ$-cohomology of the $\frac{1}{16}$-BPS operators (i.e.~the set of all such physical states) 
can change between the free theory and even very small coupling. 
As a consequence, if we count these operators (with all plus signs), the answer 
at one-loop could differ drastically from the free theory, 
It was shown in~\cite{Janik:2007pm} that this is indeed the case. 
We note, however, that the issue of loop corrections in the SYM Hamiltonian is not really relevant for us here, 
because we will only consider the supersymmetric index  which is designed to be protected~\cite{Witten:1982df}. 
Thus, assuming that there are no states coming in from infinity at zero coupling, 
the exponential growth of states should be seen in the index calculated in the free SYM theory. 
The complete $\CQ$-cohomology at weak coupling was calculated in the infinite-$N$ limit (i.e., ignoring all trace relations)
in~\cite{Chang:2013fba} following earlier work of~\cite{Janik:2007pm,Grant:2008sk}.
It was shown that the single-trace $\CQ$-cohomology is in one-to-one correspondence with the 
free graviton BPS states in the dual AdS$_5$, which is indeed a much smaller number than is expected 
from the black hole. 

\vskip 0.2cm

It will be useful for us to record the value of the trace~\eqref{indtrace} over the 
infinite-$N$~$\CQ$-cohomology in SYM/BPS gravitons in AdS. 
From the AdS point of view this is straightforward and involves listing the BPS single-graviton 
operators~\cite{Gunaydin:1984fk} and calculating the index~\cite{Kinney:2005ej}. 
In this manner one obtains 
\be \label{igrav}
\ig (\x)  \= \frac{3 \, \x^2}{1-\x^2} - \frac{2\, \x^3}{1-\x^3} \, .
\ee
From the SYM point of view one obtains the same answer 
by calculating the index over gauge-invariant single-trace operators at infinite~$N$. We can check this  
by specializing the partition function Equation (4.6) of~\cite{Chang:2013fba}  to our index calculation.
These operators also exist at finite~$N$ and we call them ``graviton operators in the SYM" and 
we shall use the notation~$\ig(\x)$ to refer to their index in the following presentation.

\vskip 0.2cm

The recent progress in this problem relies on a careful study of the index~\eqref{Uact}. 
The Hamiltonian calculation to reach this integral expression is as follows~\cite{Romelsberger:2005eg, Kinney:2005ej}. 
We use the language of operators. 
One first calculates the ``single-letter index", namely the trace~\eqref{indtrace} taken over 
all operators made up of the elementary fields of the theory and derivatives. 
In this calculation one has to be careful about subtracting the constraints 
arising from the equations of motion. The only fields (or constraints) contributing to the index 
are those which are annihilated by the supercharge~$\CQ$. The result of this calculation is
\be \label{defis}
\is (\x) \= \frac{3\, \x^2 -3 \, \x^4- 2\, \x^3 + 2\, \x^6 }{(1-\x^3)^2} 
\=  3\x^2 - 2\x^3 - 3\x^4 + 6\x^5 - 2\x^6 - 6\x^7 + 9\x^8 - 2\x^9 \+ \cdots \,.
\ee
The denominator in this formula, and the consequent infinite series on the right-hand side, 
reflects the two independent supersymmetric spacetime derivatives.  
The single-letter operators counted by~\eqref{defis} are not gauge-invariant and, in order to count 
physical states, one needs to project to the gauge-invariant subspace.
This is done by integrating over the gauge group using the Haar measure.
In this manner we reach the matrix model~\eqref{Uact}. 
For future use we note that the single-letter index can be written as follows,
\be \label{isagain}
\is (\x) \=  1-\frac{(1-\x^2)^3}{(1-\x^3)^2} \,.
\ee

\vskip 0.2cm

It was implicit in the above discussion that the single graviton operators (single-trace operators at infinite~$N$) 
and the single-letter index at finite~$N$ are building blocks from which we calculate multi-graviton and multi-trace indices. 
This is achieved by the operation of plethystic exponentiation\footnote{See~\cite{Benvenuti:2006qr} for a definition 
and a nice review of plethystics in related problems.}  on the building block answers.  
At infinite~$N$ one first projects on to gauge invariant single-traces at~$N=\infty$ as in~\eqref{igrav}, 
and then exponentiates them to produce multi-traces. 
In the full microscopic formula~\eqref{Uact}, in contrast, one first exponentiates the states charged under the 
gauge group and then projects on to gauge-invariant states. The two operations do not commute and 
therefore the two calculations give rise to different answers. 
In the next couple of sections we consider the integral~\eqref{Uact} in the large-charge and small-charge limits, 
respectively, and show that it interpolates between the multi-graviton answer and the black hole answer.

\section{Large charge operators form the black hole \label{sec:large}}

In this section we first briefly review the analytic results for the entropy, and then compare them 
with numerical results.

\vskip 0.1cm

In the large-$N$ saddle-point analysis of~\cite{Cabo-Bizet:2019eaf}, the integral is first written in a standard manner as an integral 
over the~$N$ variables~$u_i$, where the eigenvalues of the matrix are~$\rme^{2 \pi \i u_i}$. The fact that the 
integrand of~\eqref{Uact} only contains the adjoint representation implies that the action of the eigenvalues 
does not have a single-particle potential and is a pure two-particle interaction. Thus the solutions are expected to have  
equally-spaced eigenvalues. The starting point of~\cite{Cabo-Bizet:2019eaf} was the observation 
that the saddle-point configurations can have the eigenvalues spreading out into the complex~$u_i$ plane. 

In order to find saddle-points, one rewrites the integrand of~\eqref{Uact} in terms of the 
Bloch-Wigner elliptic dilogarithm~\cite{Bloch} which is doubly periodic in the complex plane with periods~$1$ and~$\t$
(recall from~Footnote~\ref{qtaurel} that~$\x=\rme^{2\pi \i \t/3}$), 
and therefore there is a saddle for every periodic configuration of a string of eigenvalues. 
Every pair of integers~$(m,n)$ with~$\text{gcd}(m,n)=1$ gives rise to an independent saddle,
which is interpreted as the string of eigenvalues with winding numbers~$(m,n)$ around the two
cycles of the torus~$\IC/(\IZ \, \t+~\IZ)$.
The effective action~$S_\text{eff}(\t)$ at each saddle can be calculated very simply using the 
double-Fourier expansion of the Bloch-Wigner elliptic dilogarithm~\cite{ZagierOnBloch}. 
Comparing the various saddles gives the phase structure of the theory as a function of~$\t$. 
The black hole corresponds to the saddle~$(1,0)$, and dominates the phase diagram 
near~$\t \to 0$. Up to a~$\tau$-independent imaginary term which is not relevant for us here, 
it has the following effective action~\cite{Cabo-Bizet:2019eaf},
\be \label{Seff10}
\Seff(1,0;\t)    
\=\frac{\pi \i\,  N^2\, (2\tau\, +\,1)^3}{27\, \tau ^2} \,,
\ee
which agrees on the nose with the supergravity calculation of the regularized on-shell action of the 
AdS$_5$ BH~\cite{Cabo-Bizet:2018ehj}. The semiclassical (large-$N$) entropy of the BH is given by the real part of the 
Legendre transform of this effective action\footnote{The entropy function~$\CE(\t,n)$ differs from the 
one presented in~[\cite{Cabo-Bizet:2019eaf}, \S 4.2] by a $\t$-independent imaginary term 
which does not affect the final entropy. We have also factored out an overall factor of~$N^2$.} (with~$j=n/N^2$), 
\be \label{SBHlargeN}
\SBH \=  \Re \, \int \, d\t \, \exp \bigl( N^2 \, \CE(\t) \bigr) \,,
\qquad \CE(\t) \=  - \frac{2 \pi \i \, \t}{3} \, j - 
\frac{\pi \i\,  (2\tau\, +\,1)^3}{27\, \tau ^2}   \,.
\ee
Denoting by~$\t_*=\t_*(j)$ the solution of the extremization equation i.e.~$\CE'(\t_*)=0$, 
we obtain, at leading order in the large-$N$ saddle-point expansion, 
\be  \label{SBHsfunction}
\SBH (N,n) \=  N^2 \, s(n/N^2) \,, \qquad s(j) \= \Re \, \CE(\t_*(j))  \,.
\ee
Since~$\CE'$ is a cubic polynomial in~$1/\t$, the solution~$\t_*$ can be found in radicals. 
The asymptotic expression~\eqref{sSBH} is obtained by further expanding in large~$j$.

\vskip 0.2cm

The microscopic index~$d_N(n)$, for fixed~$N$, is found by inverting~\eqref{indtrace}.
The Cardy-like limit~$n \to \infty$ is dominated by the~$\t \to 0$ behavior of the index function~$\CI_N$. 
This limit has been calculated in two ways---by using an estimate for the Cardy-like limit of the 
elliptic gamma function~\cite{Rains:2006dfy,Ardehali:2015bla}, as well as by extending the 
elliptic deformation method to finite~$N$~[\cite{Cabo-Bizet:2019eaf}, \S 4.5]. One finds that
\be \label{CardyIN}
\log \CI_N(\t \to 0) \= -\Seff(1,0;\t\to 0) \= - 
\frac{\pi \i N^2}{27} \Bigl(\frac{1}{\t^2} + \frac{6}{\t} \Bigr) +O(1) \,,
\ee
implying that there are no~$1/N$ corrections to the large-$N$ answer~\eqref{SBHlargeN} in the Cardy-like limit.
The asymptotics of~$\log |d_N(n)|$ as~$n \to \infty$ is given by the real part of the 
Legendre transform of the expression~\eqref{CardyIN}. 
One point worth noting here is that the behavior of the effective action in the Cardy-like limit~$\t \to 0$ goes 
as~$1/\t^2$---and not~$1/\t$ as for classical modular forms transforming under~$SL_2(\IZ)$. 
There does seem to be a symmetry underlying the index, but it is more subtle than ordinary modular symmetry, 
the effective action is a period function rather than a modular form~\cite{Cabo-Bizet:2019eaf}.

\vskip 0.2cm

\begin{figure}[h]\centering
\includegraphics[height=4.6cm]{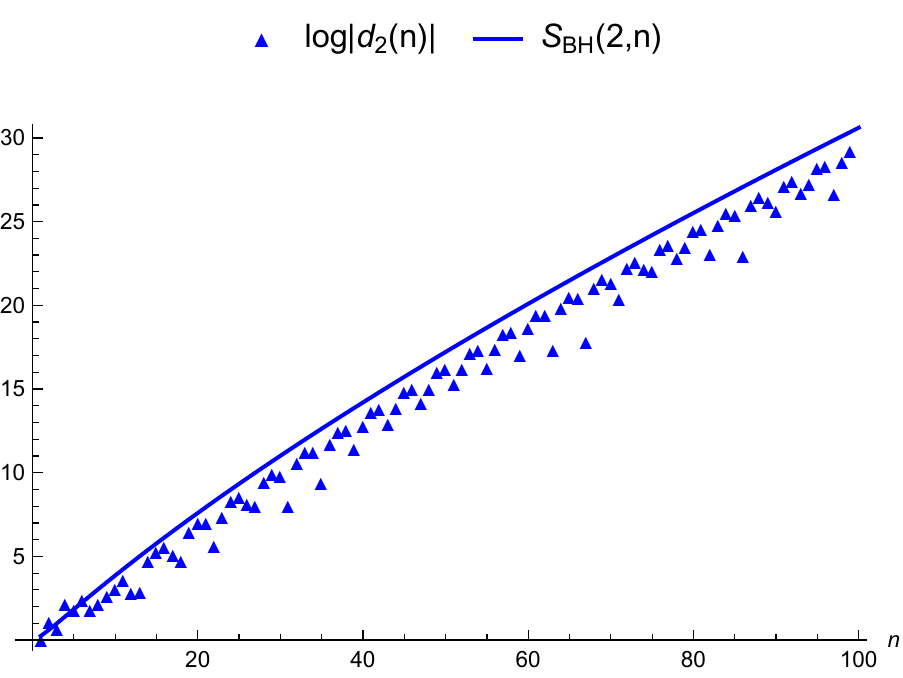}
\hspace{1cm}
\includegraphics[height=4.8cm]{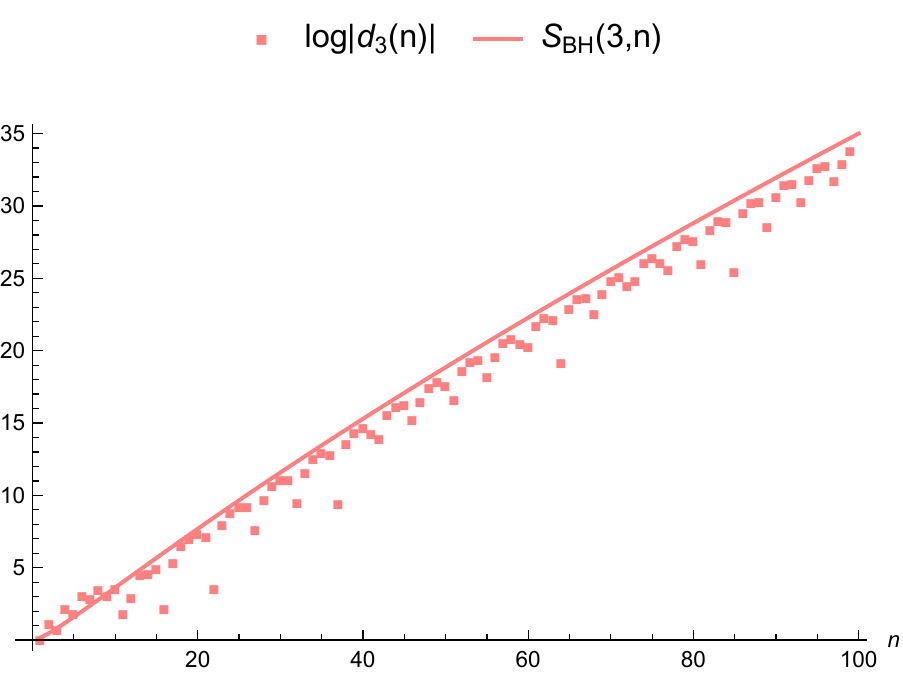}

\vspace{0.6cm}

\includegraphics[height=4.6cm]{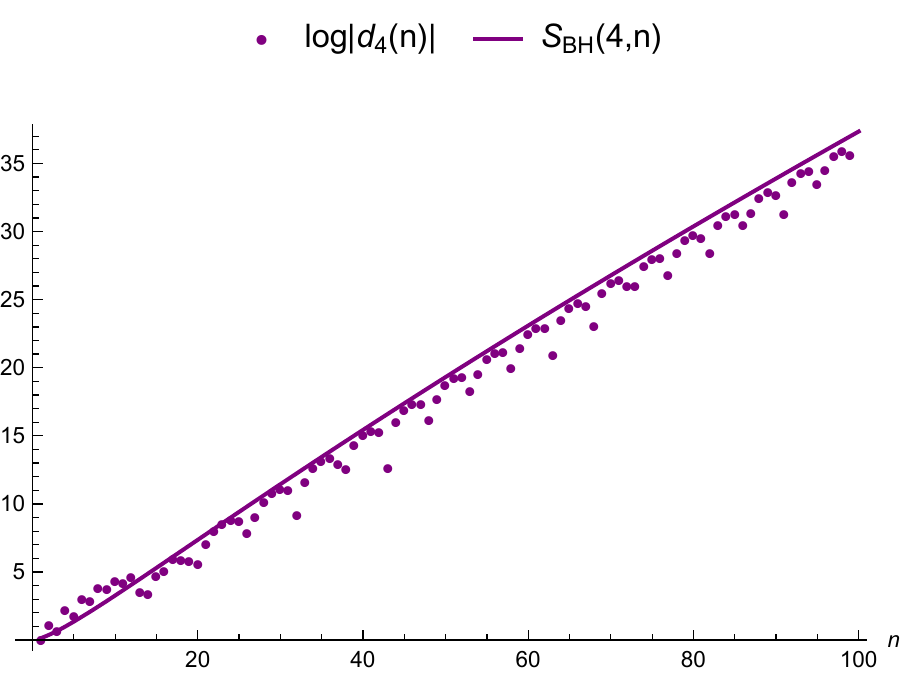}
\hspace{1cm}
\includegraphics[height=4.8cm]{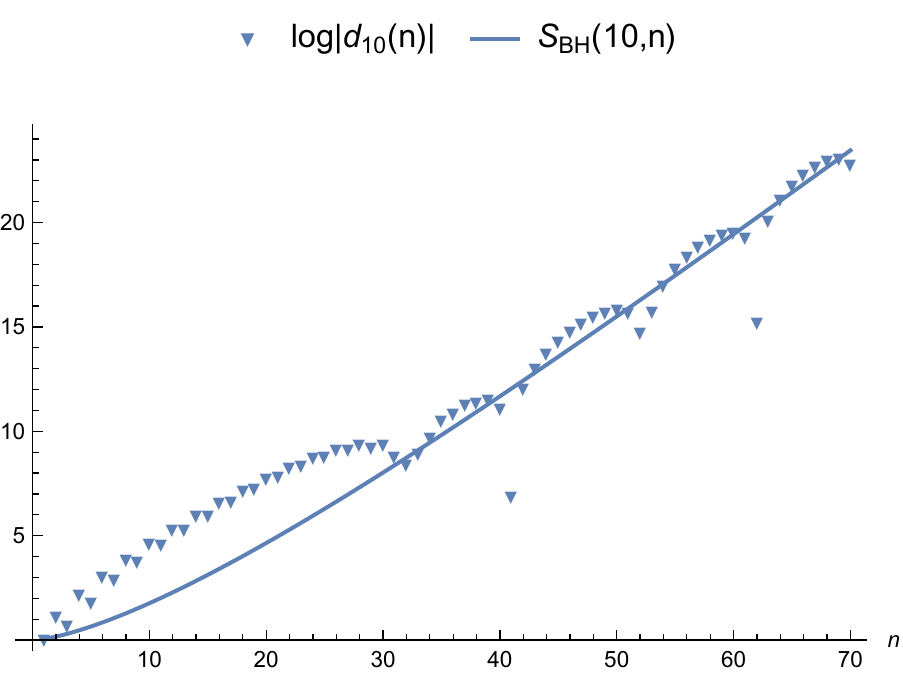}
\caption{The comparison between the microscopic entropy~$\log |d_N(n)|$ and the 
  BH entropy~$\SBH(N,n)$ for~$N=2,3,4,10$.
  The plot for~$N=10$ is zoomed in slightly, and clearly shows the initial deviation from the BH curve.
  The microscopic data for~$N=2,3,4$ are computed directly using the gamma function representation.
  The time taken to calculate~$d_N$ increases rapidly with~$N$---after initialization, and putting a cutoff 
  at~$n=100$, it took 5 ms for~$N=2$ and 26 min for~$N=4$. 
  (All these calculations were performed using PARI/GP \cite{PARI2} on a MacBookPro 2017.) 
  For higher values of~$N$ we use the formula~\eqref{INpartformula}. In this method, the computational bottleneck
  is to produce the characters of the permutation group~$S_d$, which leads to the charge cutoff~$n \le 2d$.
  The time taken to calculate the character tables from d=1 to 20 was 4 seconds 
  while the final case dealt with here, namely~$d=35$ alone took 20 hours. 
  (All the character tables were computed using GAP~\cite{GAP4}.)
  Having obtained the characters, calculating the coefficients~$d_N$ is quite fast, 
  e.g.~the case~$N=10$, $n \le 70$ took 14 min using PARI/GP.
} 
  \label{MicMaccomparison}
\end{figure}

In Figure~\ref{MicMaccomparison} we present the comparison between the microscopic integers~$\log |d_N(n)|$ 
and the expression~\eqref{SBHlargeN} for the BH entropy for~$N=2,3,4,10$. 
There are many points to note here.
Firstly we find agreement as~$n \to \infty$, as expected from the above discussion. 
For small charges, until~$2N$, the microscopic degeneracies deviate from the BH curve. 
As explained analytically in Section~\ref{sec:small}, they follow the graviton curve for these small charges.

\begin{figure}[h]\centering
\includegraphics[height=4.5cm]{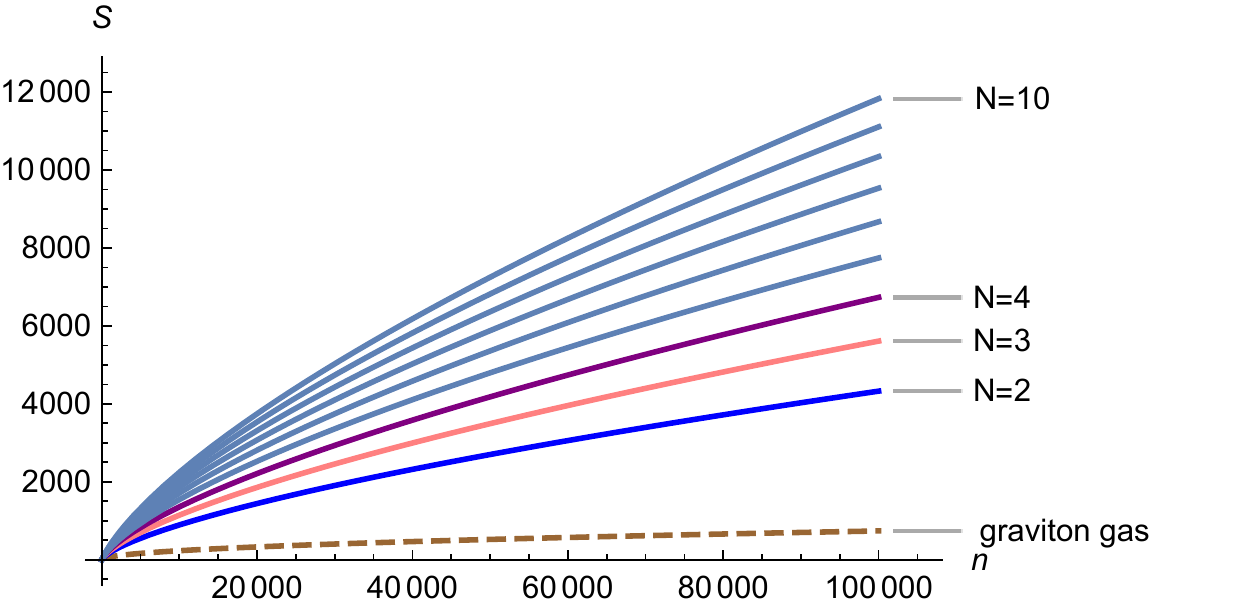}
  \caption{The gravitational black hole entropy~$\SBH=N^2 s(j) = a_2(Nn)^{2/3} + O(n^{1/3})$ 
  for~$N=2,3,\dots,10$ and the logarithm of the index of the graviton gas (dashed line).} 
  \label{SBH20}
\end{figure}
\begin{figure}[h]\centering
\includegraphics[height=4.2cm]{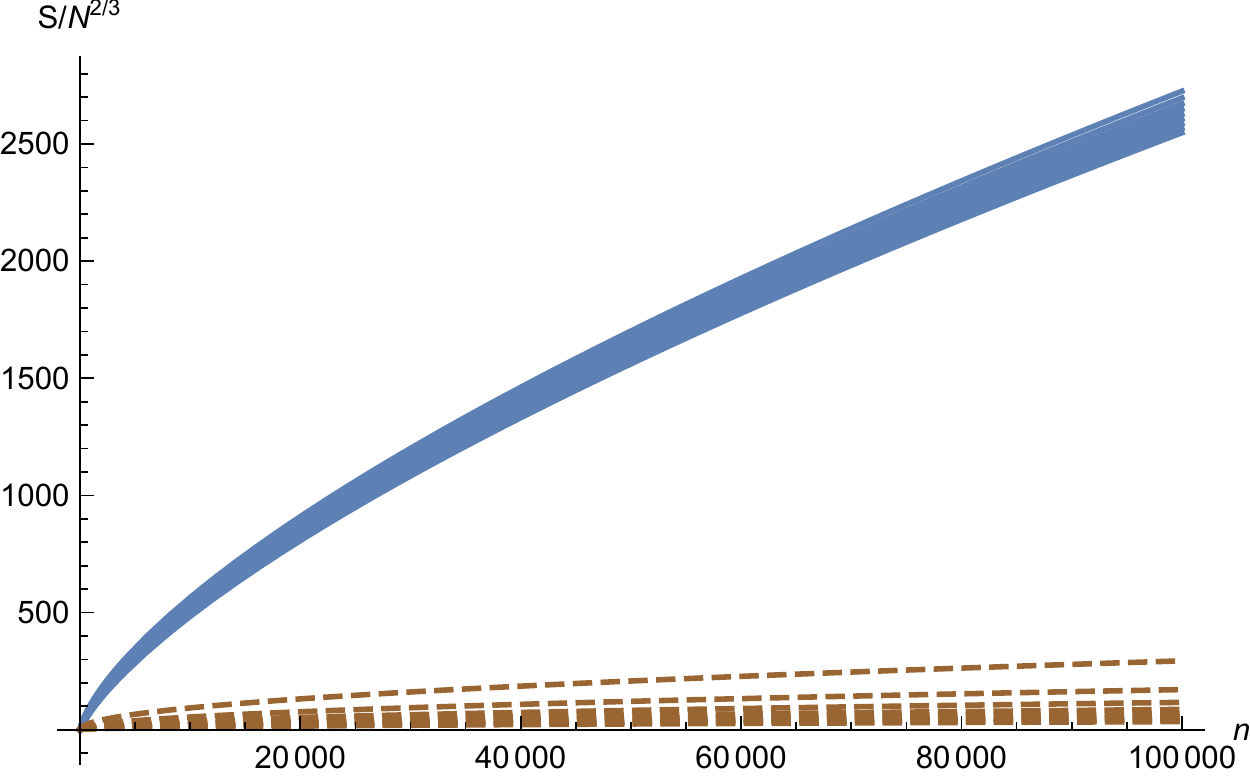}
\hspace{0.5cm}
  \includegraphics[height=4.2cm]{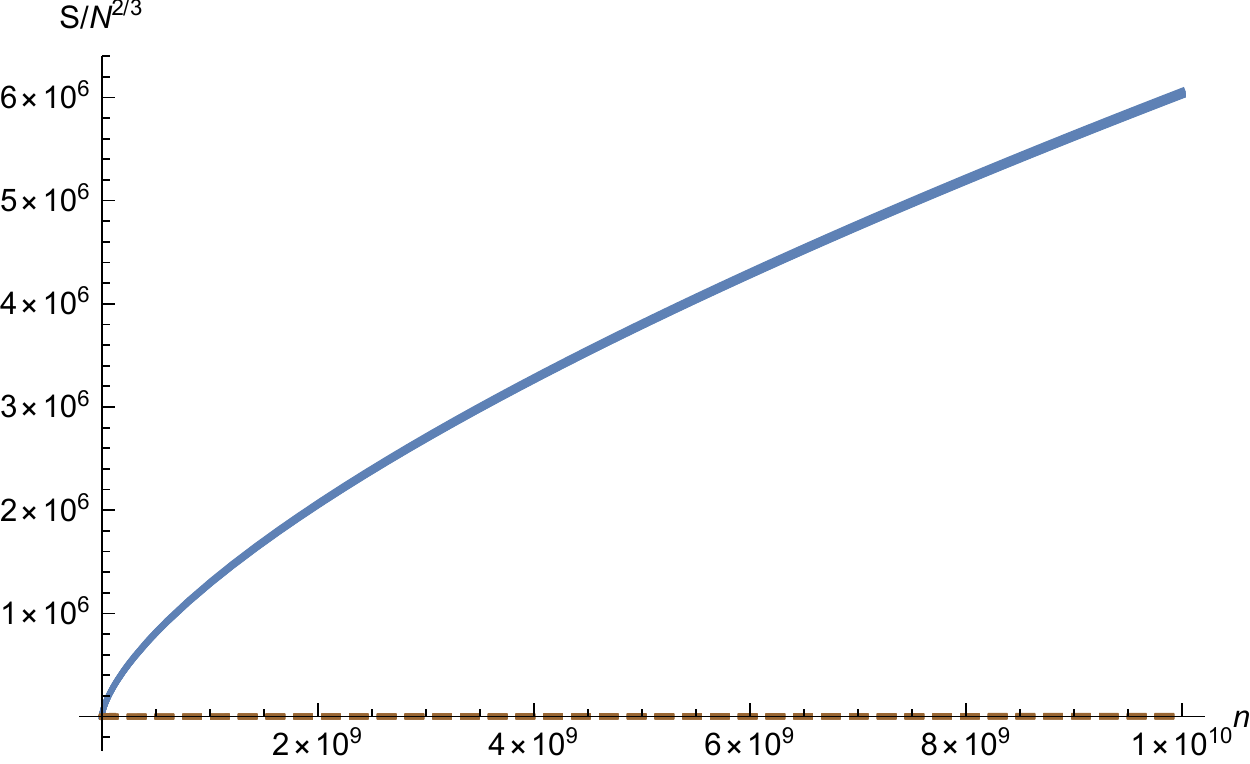}
  \caption{The blue (solid) lines are~$\SBH(N,n)/N^{2/3}$ as a function of~$n$ (left: $N = 2, \dots 10$, 
  right: $N=2, \dots 100$). 
  The brown (dashed) lines are~$\log d_\text{grav}(n)/N^{2/3}$, as a function of~$n$. 
  The blue lines converge to the uppermost line as~$N \to \infty$ and 
   the brown lines converge to zero as~$N \to \infty$. The convergence is seen more clearly for larger~$N$ on the right.}
  \label{fig:conv}
\end{figure}

After charge~$2N$, the degeneracies latch on to the BH curve very soon (within~$O(N)$), and 
exhibit regular bumps of size~$N$, which would be interesting to explain. 
(The bumps are smoothed out in the analytic formulas plotted in Figure~\ref{SBH20}.)
In particular, the good agreement of the microscopics with the BH for small~$j=n/N^2$ and small~$N$ is remarkable.
It is not a consequence of any known controllable analytic formula, 
and suggests that the~$O(1)$ correction to the Cardy-like limit in~\eqref{CardyIN} is given exactly by the large-$N$
results~\eqref{SBHlargeN}, while a priori one could have added any other function of~$\tau$ that vanishes as~$\t \to 0$.
We caution the reader that our numerics are not precise enough to declare such a conclusion, but they do prompt
the conjecture that the simple expression~\eqref{SBHlargeN} essentially governs the perturbative entropy even at finite~$N$.

\section{Small charge operators are gravitons \label{sec:small}}

In this section we turn to small~$n$ and compare the values of the index of gauge-invariant operators in SYM 
with the index in the Hilbert space of multi-graviton operators. 

\vskip 0.1cm

The index function of multi-graviton states is, by definition, the plethystic exponential of the single-graviton
index~$\ig$, given in~\eqref{igrav},
\be \label{multigrav}
\CI_{\text{multi-grav}} (\x) \= 
\sum_n \, d_\text{grav}(n)\, \x^n \defeq \exp\Bigl( \, \sum_{k=1}^\infty\, \frac{1}{k} \, \ig(x^k)  \Bigr)
\= \prod_{n=1}^\infty \, \frac{(1-\x^{3n})^2}{(1-\x^{2n})^3} \= \frac{\eta(\t)^2}{\eta(\frac{2\t}{3})^3}\,.
\ee
We see from this expression that the multi-graviton index is equivalent to a gas of three real bosonic 
oscillators of frequencies~$2n$ and one complex fermionic oscillator of frequencies~$3n$, $n=1,2,\dots$. 
Using the standard modular properties of the
Dedekind~$\eta$ function to estimate the growth of states, we obtain 
\be
\log d_\text{grav}(n)  \; \stackrel{n\to \infty}{\longrightarrow} \; \frac{\pi}{3} \sqrt{5n} \,,
\ee
which is equivalent to an effective central charge of~$\frac56$.
In Figures~\ref{SBH20}, \ref{fig:conv} we show the growth of~$\log d_\text{grav}(n)$ 
in comparison to the BH entropy.

\begin{figure}[h]\centering
  \includegraphics[height=5.2cm]{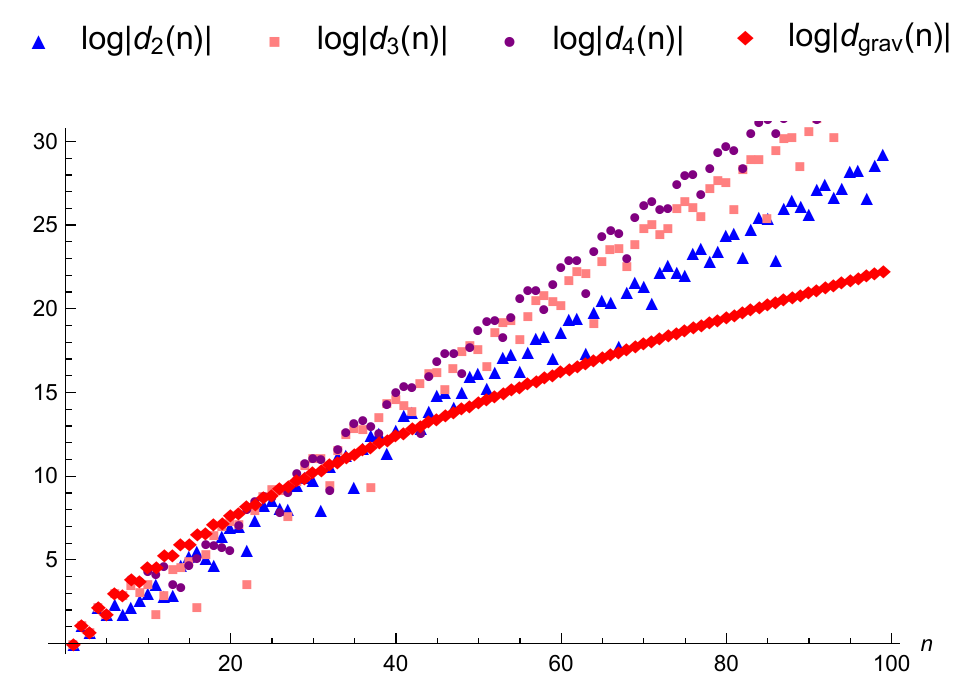}
  \caption{Microscopic data of~$d_N(n)$ for $N= 2,3,4$, and of~$d_\text{grav}$.}
  \label{Fig:N=234grav}
\end{figure}

We now discuss the gauge theory index over the whole range of charges. 
The pattern is as follows. For~$\frac12 n < N+1$ the gauge theory index~$d_N$
and the multi-graviton index~$d_\text{grav}$ agree exactly, as illustrated in Table~\ref{tab:gravdncomp},
and as we prove below.
As we increase the charge~$n$,~$d_N$ falls behind for a small interval before picking up 
and dominating~$d_\text{grav}(n)$ at large~$n$, as shown in Figure~\ref{Fig:N=234grav}. 
For very large charges, $d_N$ agrees with the BH partition function as discussed in the previous section.
It seems to be important for these observations at small charges that we 
are considering~$U(N)$ and not~$SU(N)$.\footnote{That AdS$_5$/CFT$_4$ should be 
formulated in terms of~$U(N)$ has been argued for previously in~\cite{Maldacena:2001ss,Belov:2004ht}.}
Indeed, the projection onto the gauge-invariant subspace of the~$k=1$ term in the exponential in~\eqref{Uact} 
for~$SU(N)$  theories is empty.

We recall from~\eqref{indtrace},~\eqref{Uact} that the gauge theory index is
\be \label{dN1}
\CI_N(\x) \= \sum_n \, d_N(n)\, \x^n \= \int \, DU\, \exp \biggl( \; \sum_{j=1}^\infty \, \frac{1}{j} \, 
\is(\x^j) \, \Tr \, U^j \, \Tr \, (U^\dagger)^j \, \biggr) \,.
\ee
Using the relation~\eqref{isagain}, we write the multi-graviton index~\eqref{multigrav} as
\be\label{dgrav1}
\CI_{\text{multi-grav}} (\x) \= \sum_n \, d_\text{grav}(n)\, \x^n \= \prod_{k=1}^\infty \frac{1}{1-\is(\x^k)} \,.
\ee
Our goal now is to prove that~$d_N(n)$ agrees with~$d_\text{grav}(n)$ when~$n \le  2N +1$. 
Expanding both the expressions~\eqref{dN1}, \eqref{dgrav1} in terms of products of~$\is(\x^k)$
over different~$k$, and recalling from~\eqref{defis} that the index~$\is(\x)$ starts with the power~$\x^2$
we see that this agreement is equivalent to the following assertion,
\be \label{assertion}
\int \, DU\,  \prod_{j=1}^m \, \frac{1}{k_j ! \, j^{k_j}} 
\bigl(\Tr \, U^j \, \Tr \, U^{\dagger j} \bigr)^{k_j}  \=1 \,, \qquad \sum_{j=1}^m \, j \, k_j \; \le \; N\,,
\ee
which we now prove using some simple concepts from representation theory of~$U(N)$.  
The ideas below have appeared in closely related contexts in~\cite{Dolan:2007rq,Dutta:2007ws}.

\begin{table}[h!]
\centering
\begin{tabular}{|c|cccccccccccccccc|} 
\hline
$n$ & 2 & 3 & 4 & 5 & 6 & 7 & 8 & 9 & 10 & 11 & 12 & 13 & 14 & 15  & 16 & 17 \\ [0.5ex] 
 \hline  
 \hline  
 $d_2$ & \blue{3}& \blue{-2}& \blue{9}& \blue{-6}& {11}& {-6}& {9}& {14}& {-21}& {36}& {-17}& {-18}& {114}
	& {-194}& {258}& {-168}  \\ [0.5ex] 
\hline  
 $d_3$ & \blue{3}& \blue{-2}& \blue{9}& \blue{-6}& \blue{21}& \blue{-18}& {33}& {-22}& {36}& {6}
 	& {-19}& {90}& {-99}& {138}& {-9}& {-210}\\ [0.5ex] 
 \hline  
 $d_4$ & \blue{3}& \blue{-2}& \blue{9}& \blue{-6}& \blue{21}& \blue{-18}& \blue{48}& \blue{-42}& {78}& {-66}
 	& {107}& {-36}& {30}& {114}& {-165}& {390}  \\ [0.5ex] 
\hline  
 $d_5$  & \blue{3}& \blue{-2}& \blue{9}& \blue{-6}& \blue{21}& \blue{-18}& \blue{48}& \blue{-42}& \blue{99}
 	& \blue{-96}& {172}& {-156}& {252}& {-160}& {195}& {48} \\ [0.5ex] 
\hline
 $d_6$  & \blue{3}& \blue{-2}& \blue{9}& \blue{-6}& \blue{21}& \blue{-18}& \blue{48}& \blue{-42} & \blue{99}
 	& \blue{-96}& \blue{200}& \blue{-198}& {345}& {-340}& {540}& {-426} \\ [0.5ex] 
\hline
\hline
$d_\text{grav}$  & \blue{3}& \blue{-2}& \blue{9}& \blue{-6}& \blue{21}& \blue{-18}& \blue{48}& \blue{-42}& \blue{99}& \blue{-96}& \blue{200}& \blue{-198}& \blue{381}& \blue{-396}& \blue{711}& \blue{-750}  \\ [0.5ex] 
\hline
\end{tabular}
\caption{The $U(N)$ SYM index $d_N(n)$ equals $d_\text{grav}(n)$ for~$\frac12 n < N+1$, and then
starts to differ. More data is shown in Table~\ref{tab:gravdncomp2} at the end of the paper.}
\label{tab:gravdncomp}
\end{table}

The basic idea is to expand the traces of powers of the gauge field in terms of the 
$U(N)$ group characters, which is precisely the content of the Frobenius character formula~\cite{FultonHarris,Teleman}. 
Recall that the representations of~$U(n)$ and those of the symmetric group~$S_n$ are both 
labelled by partitions~$\bl$ of~$n$. We denote the corresponding characters as~$\wt\chi_{\bl}$
and~$\chi^{\bl}$, respectively. 
Now note that the gauge theory operator in~\eqref{assertion}
is uniquely associated with a cycle shape~$P$ through the following bijection,
\be  
P \;\equiv \; \prod_{j=1}^m (j)^{k_j} \; \; \longleftrightarrow \; \;
\prod_{j=1}^m \, \bigl(\Tr \, U^j \bigr)^{k_j} \; \equiv \; \CO_P (U)\,.
\ee
We think of~$P$ as a partition of the integer~$|P| \coloneqq \sum_{j=1}^m  j  k_j$,
labelling a conjugacy class in~$S_{|P|}$. Here~$|P|$ is called the weight of
the partition~$P$, and the number~$\ell(P)\coloneqq\sum_{j=1}^m  k_j$ is called 
the length or the number of parts of the partition~$P$. 
The Frobenius formula for~$U(N)$ states that 
\be \label{Frobfor}
\CO_P (U) \= \sum_{\bl  \;\vdash \; |P| \atop \ell(\bl)\le N} \, \wt \chi_{\bl}(U) \, \chi^{\bl}(P) \,,
\ee
where the notation~$\bl \vdash  n$ means~$\bl$ is a partition of~$n$,  
so that the sum runs over all partitions of~$|P|$ with at most~$N$ parts. 

Using the formula~\eqref{Frobfor} for~$\CO_P(U)$ and  
using the first orthogonality relation of the group characters of~$U(N)$, i.e.,
\be
\int \, DU\,   \wt \chi_{\bl}(U) \, \wt \chi_{\bl'}(U^\dagger)  \= \delta_{\bl \bl'} \,,
\ee
we obtain
\be \label{upupbarint}
\int \, DU\, \CO_P (U) \, \CO_P(U^\dagger)  
\= \sum_{\bl  \;\vdash \; |P| \atop \ell(\bl)\le N}  \, \chi^{\bl}(P) \; \overline{\chi^{\bl}(P)} 
\= \sum_{\bl  \;\vdash \; |P| \atop \ell(\bl)\le N}  \, \chi^{\bl}(P)^2 \,. 
\ee 
Here the second equality is a consequence of the fact that the characters of the symmetric group
are real (in fact, integers).

When~$|P| \le N$, any partition of~$|P|$ cannot have more than~$N$ parts, so that 
the sum over~$\bl$ on the right-hand side of~\eqref{upupbarint} runs over~\emph{all} partitions of~$|P|$.
In this case we can use the second orthogonality relation of the characters of the symmetric group, i.e.,
\be \label{Orth2}
\sum_{\bl  \;\vdash \; |P|} \, \chi^{\bl}(P)^2  \= \prod_{j=1}^m k_j ! \, j^{k_j} \; \equiv \; z_P \,.
\ee
Upon putting the equations~\eqref{upupbarint}, \eqref{Orth2} together, we obtain the assertion~\eqref{assertion}. 
It is important in this argument that~$|P| \le N$, this condition guarantees that the power of~$\x$ 
in the index~\eqref{dN1} is less than~$2(N+1)$. As long as this holds, the coefficient~$d_N$ is independent of~$N$ and agrees 
with~$d_\text{grav}$---which is manifestly independent of~$N$.

In fact the Frobenius relation can also be used to write down an explicit formula  
for the index~\eqref{dN1}. Upon expanding the exponential in~\eqref{dN1} and using 
the formula~\eqref{upupbarint} we obtain
\be \label{INpartformula}
\begin{split}
\CI_N(\x) & \= \int \, DU\,  \sum_{P} \frac{1}{z_P} \, \is(\x)_P \; \CO_P (U) \, \CO_P(U^\dagger)  \\
& \= \sum_{P} \, \is(\x)_P \; 
 \frac{1}{z_P} \, \sum_{\bl  \;\vdash \; |P| \atop \ell(\bl)\le N}  \, \chi^{\bl}(P)^2  \,,
 \end{split}
\ee
where the first sum (over~$P$) runs over all partitions, and we have introduced the notation
\be
f(\x)_P \defeq \prod_{j=1}^m f(\x^j)^{k_j} \,.
\ee
When~$N \to \infty$ at fixed charges we see, from~\eqref{Orth2} and \eqref{INpartformula}, that 
\be
\CI_{N=\infty} (\x) \= \sum_{P} \is(\x)_P \= \CI_{\text{multi-grav}} (\x) \,,
\ee
as consistent with our observations above.
The formula~\eqref{INpartformula} can be used to efficiently produce numerical calculations 
for the degeneracies~$d_N$ as shown in the previous section.

\vskip 0.2cm

We end with some brief comments. The matrix model that we study captures the 
index of~$\frac{1}{16}$-BPS states, which is protected by supersymmetry. It manages to capture
the low growth behavior of supergravitons at low energies and transitions to the exponential BH growth 
at high energies. One question of substantial interest to understand the details of 
the~$\frac{1}{16}$-BPS operators at large charges. The idea of the sum over random partitions
discussed here, and the naturally related free fermions, have been discussed in this  
context~\cite{Berenstein:2004kk, Berkooz:2006wc}, and it would be interesting to try to push these
ideas further.

On a slightly philosophical note, 
we propose that the equations~\eqref{multigrav},~\eqref{dN1},~\eqref{dgrav1} should be 
thought of as a producing a \emph{black hole transform} of the single-graviton index~\eqref{igrav}. 
The input single-graviton index can be calculated using only the global symmetries of the AdS theory, and 
the transform~$d_N$ informs us about large BH solutions.
Using the relation of the single-letter trace to the graviton index, we can also think of this transform 
as a holographic relation between the gauge-theory single-letter trace and the black hole. 

Another interesting question is if there is a theory on the gravitational side which directly captures 
the dynamics of these~$\frac{1}{16}$-BPS states. 
As a first shot we might imagine some kind of topological theory, but the appearance of black 
holes---which we clearly see in the matrix model here---is not seen in the usual topological versions of 
AdS/CFT dualities~\cite{Gopakumar:1998ki,Ooguri:2002gx},
This should be related to the fact that in this note we look at energies which scale at the same 
rate as~$N^2$ while the usual treatments took~$N \to \infty$ strictly. 

Finally, the main mathematical context for the ideas used here to probe the Hawking-Page transition is 
the Frobenius-Schur duality, which relates the representations of~$U(N)$ and those of the symmetric group. 
This idea has been used fruitfully in the past to relate matrix models appearing in gauge theories to string 
theories~\cite{Gross:1993hu, Douglas:1993iia, Cordes:1994fc}. 
It would be interesting to see if it can be used to construct some type of topological 
theory on AdS$_5$, defined at finite~$N$, that is dual to states captured by the matrix model studied here.

\section*{Acknowledgements}
We would like to thank Prarit Agarwal, Dionysios Anninos, Alejandro Cabo-Bizet, 
Nikolay Gromov, Amihay Hanany, Juan Maldacena, 
Mukund Rangamani, and Don Zagier for useful discussions.
This work is supported by the ERC Consolidator Grant N.~681908, ``Quantum black holes: A microscopic 
window into the microstructure of gravity'', and by the STFC grant ST/P000258/1.

\begin{table}[h!]
\centering
\begin{tabular}{|c||c|c|c|c|c|c||c|c|c|c|} 
\hline
$n$ & $d_2$ &  $d_3$ &  $d_4$ &  $d_5$ &  $d_6$ &  $d_7$ &  $d_\text{grav}$ \\ [0.5ex] 
 \hline  
 \hline  
 14 &114 &-99 &30 &252 &345 &381 &381 \\ [0.5ex]
\hline
15 &-194 &138 &114 &-160 &-340 &-396 &-396 \\ [0.5ex]
\hline
16 &258 &-9 &-165 &195 &540 &666 &711 \\ [0.5ex]
\hline
17 &-168 &-210 &390 &48 &-426 &-678 &-750 \\ [0.5ex]
\hline
18 &-112 &672 &-366 &-127 &564 &1059 &1278 \\ [0.5ex]
\hline
19 &630 &-1116 &330 &612 &-234 &-960 &-1386 \\ [0.5ex]
\hline
20 &-1089 &1554 &276 &-783 &189 &1311 &2256 \\ [0.5ex]
\hline
21 &1130 &-1270 &-1212 &1258 &636 &-900 &-2472 \\ [0.5ex]
\hline
22 &-273 &-36 &3081 &-948 &-1026 &1017 &3879 \\ [0.5ex]
\hline
23 &-1632 &2898 &-4986 &450 &2262 &150 &-4320 \\ [0.5ex]
\hline
24 &4104 &-6705 &6924 &1921 &-2583 &-678 &6564 \\ [0.5ex]
\hline
25 &-5364 &10224 &-6654 &-5430 &3438 &2910 &-7362 \\ [0.5ex]
\hline
26 &3426 &-9918 &2616 &11793 &-1851 &-4050 &10890 \\ [0.5ex]
\hline
27 &3152 &2018 &8528 &-18812 &-794 &7012 &-12338 \\ [0.5ex]
\hline
28 &-13233 &16470 &-26571 &26379 &8757 &-7272 &17820 \\ [0.5ex]
\hline
29 &21336 &-42918 &49800 &-27750 &-20460 &7884 &-20286 \\ [0.5ex]
\hline
30 &-18319 &66906 &-67651 &17809 &40398 &-1755 &28707 \\ [0.5ex]
\hline
31 &-2994 &-66006 &63096 &15648 &-63054 &-8418 &-32886 \\ [0.5ex]
\hline
32 &40752 &13566 &-9678 &-78324 &88401 &32997 &45696 \\ [0.5ex]
\hline
33 &-76884 &106404 &-112980 &175030 &-99388 &-68454 &-52512 \\ [0.5ex]
\hline
34 &78012 &-273204 &307098 &-285576 &80856 &125910 &71811 \\ [0.5ex]
\hline
35 &-11808 &407442 &-522066 &366024 &4680 &-193416 &-82848 \\ [0.5ex]
\hline
36 &-121384 &-364710 &634029 &-323807 &-184576 &270875 &111678 \\ [0.5ex]
\hline
37 &262206 &-12024 &-436260 &38856 &494910 &-317544 &-129096 \\ [0.5ex]
\hline
38 &-293145 &778272 &-296460 &624894 &-920943 &295080 &171810 \\ [0.5ex]
\hline
39 &91904 &-1731542 &1682020 &-1718016 &1392360 &-105334 &-199080 \\ [0.5ex]
\hline
40 &359775 &2300499 &-3497613 &3094992 &-1690101 &-343179 &261900 \\ [0.5ex]
\hline
41 &-867906 &-1611774 &4937946 &-4226862 &1451568 &1189164 &-303810 \\ [0.5ex]
\hline
42 &1026540 &-1093848 &-4501122 &4098270 &-114147 &-2486379 &395538 \\ [0.5ex]
\hline
43 &-404454 &5702562 &304512 &-1210728 &-2931498 &4198644 &-459450 \\ [0.5ex]
\hline
44 &-1086312 &-10400586 &8971113 &-5968935 &8129358 &-5955660 &592512 \\ [0.5ex]
\hline
45 &2815744 &11407626 &-22380734 &18061488 &-15183836 &6982824 &-688608 \\ [0.5ex]
\hline
46 &-3415932 &-4086693 &34738953 &-33152565 &22398435 &-5795586 &880407 \\ [0.5ex]
\hline
47 &1436112 &-13996782 &-35553996 &44941584 &-25748382 &280512 &-1023840 \\ [0.5ex]
\hline
48 &3403791 &38712766 &10888602 &-41448422 &18439724 &12204848 &1298684 \\ [0.5ex]
\hline
49 &-9007578 &-56127654 &49956294 &6241896 &8645112 &-33852894 &-1510380 \\ [0.5ex]
\hline
50 &10895604 &44316099 &-142303191 &75761478 &-64166661 &64922268 &1901991 \\ [0.5ex]
\hline
\end{tabular}
\end{table}

\begin{table}[h!]
\centering
\begin{tabular}{|c||c|c|c||c|c||c|c|c|c|c|} 
\hline
$n$ & $d_2$ &  $d_3$ &  $d_4$ &    $d_\text{grav}$ \\ [0.5ex] 
 \hline  
 \hline  
51 &-4420644 &16085226 &231744000 &-2212332 \\ [0.5ex]
\hline
52 &-11068260 &-122617179 &-246464136 &2767356 \\ [0.5ex]
\hline
53 &28481682 &231054624 &90402078 &-3218130 \\ [0.5ex]
\hline
54 &-33440475 &-251544720 &309123032 &4000719 \\ [0.5ex]
\hline
55 &11822670 &80412606 &-917051802 &-4651416 \\ [0.5ex]
\hline
56 &36950502 &324099348 &1494916050 &5749536 \\ [0.5ex]
\hline
57 &-88878842 &-844286204 &-1558557796 &-6681552 \\ [0.5ex]
\hline
58 &98770059 &1147990887 &485393061 &8215209 \\ [0.5ex]
\hline
59 &-25918986 &-767030682 &2144544540 &-9542592 \\ [0.5ex]
\hline
60 &-124747447 &-628392075 &-5983505013 &11675016 \\ [0.5ex]
\hline
61 &272655942 &2808255348 &9333423798 &-13552884 \\ [0.5ex]
\hline
62 &-279580701 &-4642468821 &-9004631841 &16505106 \\ [0.5ex]
\hline
63 &35207790 &4223264234 &1231871108 &-19147932 \\ [0.5ex]
\hline
64 &419441625 &209141406 &15915475365 &23218371 \\ [0.5ex]
\hline
65 &-818211192 &-8584019040 &-38937814944 &-26916072 \\ [0.5ex]
\hline
66 &751976333 &17327115906 &55770600072 &32506014 \\ [0.5ex]
\hline
67 &54317328 &-19194283332 &-46223256036 &-37654938 \\ [0.5ex]
\hline
68 &-1386833514 &6197598675 &-10405285128 &45302448 \\ [0.5ex]
\hline
69 &2387940758 &24052600650 &118932061824 &-52435056 \\ [0.5ex]
\hline
70 &-1893048381 &-61026825105 &-247095009891 &62858988 \\ [0.5ex]
\hline
71 &-700663056 &78594793644 &311970699564 &-72696096 \\ [0.5ex]
\hline
72 &4467470232 &-43722790228 &-193686936205 &86854176 \\ [0.5ex]
\hline
73 &-6731222448 &-60628872366 &-205315072914 &-100358298 \\ [0.5ex]
\hline
74 &4333120557 &205754044713 &855723695370 &119522262 \\ [0.5ex]
\hline
75 &3746183998 &-300949636742 &-1490314195506 &-137985182 \\ [0.5ex]
\hline
76 &-13926217512 &217767461283 &1572823900839 &163839240 \\ [0.5ex]
\hline
77 &18169226454 &129914189388 &-458786822988 &-188975796 \\ [0.5ex]
\hline
78 &-8426843619 &-671070962823 &-2181976709955 &223743597 \\ [0.5ex]
\hline
79 &-15799669950 &1099745830260 &5759182587780 &-257837562 \\ [0.5ex]
\hline
80 &41774162736 &-937888762842 &-8289856609587 &304447488 \\ [0.5ex]
\hline
81 &-46405515308 &-191081792160 &6601945579040 &-350513538 \\ [0.5ex]
\hline
82 &10894454985 &2135620393074 &2245784042823 &412811109 \\ [0.5ex]
\hline
83 &58624684746 &-3884644088484 &-18254661918174 &-474836256 \\ [0.5ex]
\hline
84 &-119915881179 &3715774679244 &35440988310091 &557859048 \\ [0.5ex]
\hline
85 &110030518596 &-114903322902 &-40697268408630 &-641078046 \\ [0.5ex]
\hline
\end{tabular}
\caption{The growth of the SYM index and its large deviations from the graviton index.}
\label{tab:gravdncomp2}
\end{table}

%

\providecommand{\href}[2]{#2}\begingroup\raggedright\endgroup

\end{document}